\documentclass[aps,onecolumn,groupedaddress,amsmath,amssymb,superscriptaddress]{revtex4}

\usepackage{amsmath}
\usepackage{amssymb}
\usepackage{graphicx}
\usepackage{bm}
\usepackage{epic}
\usepackage{eepic}
\usepackage{pifont}
\usepackage[latin1]{inputenc}
\usepackage{rotating}
\usepackage{color}
\usepackage{nicefrac}
\usepackage{ulem}
\usepackage{longtable}
\usepackage{siunitx}
\usepackage[caption=false]{subfig}
\usepackage{mathtools}
\DeclarePairedDelimiter{\abs}{\lvert}{\rvert}

\begin{document}
\title{A simple algorithm to implement jump boundary \\
conditions within the lattice Boltzmann method}
\author{Badr Kaoui}
\email{badr.kaoui@utc.fr}
\affiliation{Biomechanics and Bioengineering Laboratory (UMR CNRS 7338), \\
Universit\'{e} de technologie de Compi\`{e}gne, F-60200 Compi\`{e}gne, France}
%
%
\begin{abstract}
An algorithm is proposed to implement unsteady jump boundary conditions, presenting discontinuity in physical quantities, within the lattice Boltzmann method (LBM).
This is useful to tackle problems involving mass or heat transfer through interfaces or membranes that exhibit resistance to transport.
The algorithm is simple to implement within an existing LBM based code that computes diffusion and advection of a scalar field, either temperature or solute concentration.
Analytical solutions are recovered numerically with acceptable accuracy for the limiting case of a planar membrane.
The algorithm is adapted for moving-free boundaries and adopting arbitrary geometries when combined with the immersed boundary method.
This is demonstrated by simulating controlled release of a solute from a stationary rigid particle and from a moving deformable  particle.
\end{abstract}
\maketitle
\section{Problem statement}
Many natural phenomena and technological processes involve transport of chemical species across interfaces or membranes.
For example, most of the material exchange in biological and biomedical systems take place across membranes \cite{Fournier1999,Truskey2010}.
These membranes act as a barrier to filter or control the rate of material exchange between two adjacently separated environments.
The main property that characterizes these transport functionalities, at the macroscopic scale, is the permeability or its inverse the resistance.

Modeling the effect of restricted permeability that leads to a jump in the solute concentration at the membrane is known to be technically delicate.
The resulting discontinuity represents numerical challenges specially when the membrane is moving and adopting irregular geometrical configurations.
Few works have recently been proposed to handle this issue, with applications, ranging from stationary planar interfaces or membranes to moving deformable particles.
Membranes with restrictive permeability to both, solute and solvent, or only to one of them have been considered.
For example, Wang \cite{Wang2004} and Hickson \textit{et al.} \cite{Hickson2011} have proposed finite difference schemes,
Layton \cite{Layton2006} and Jayathilake \textit{et al.} \cite{Jayathilake2010,Jayathilake2011} have used the immersed interface method \cite{Li2006}, 
Miyauchi \textit{et al.} \cite{Miyauchi2015,Miyauchi2017} have proposed a finite element discretization method,
Huang \textit{et al.} \cite{Huang2009,Gong2014} have used the immersed boundary method (IBM) \cite{Peskin2002}, 
and Yao \textit{et al.} \cite{Yao2017} have combined the IBM with a Cartesian grid embedded boundary method.
In this article, another alternative algorithm is proposed.

\begin{figure}
\centering
\includegraphics*[height=1.in, angle=0]{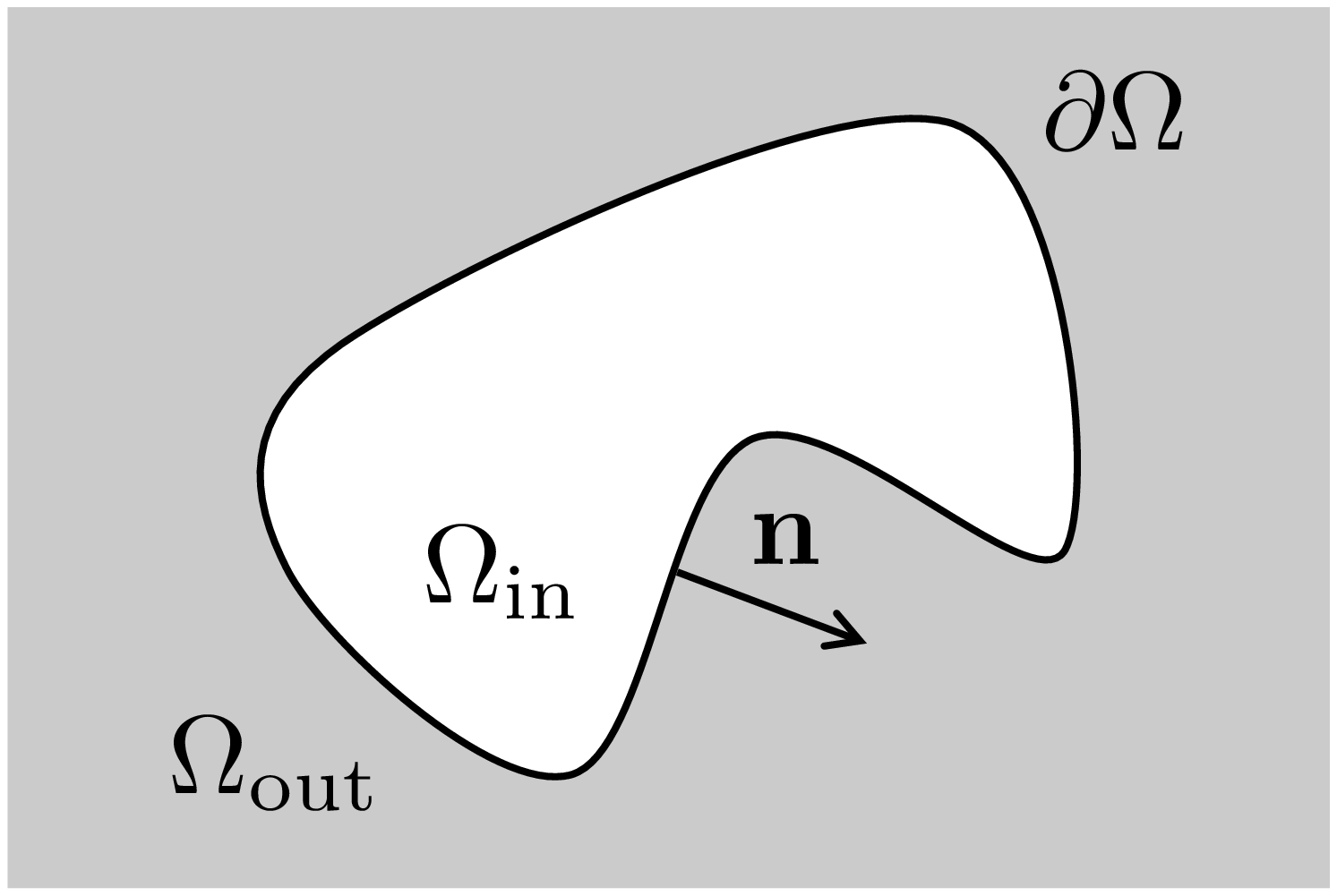}
\caption{\label{fig:mathematical1}
The computational domain $\Omega$ with its two subdomains, $\Omega _{\rm in}$ and $\Omega _{\rm out}$, separated by a closed zero-thickness membrane $\partial \Omega$. ${\bf n}$ is the unit vector normal to the membrane.
}
\end{figure}

The lattice Boltzmann method (LBM) has emerged as a modern efficient numerical method for computational fluid dynamics (CFD) \cite{Wolf-Gladrow2000,Succi2001,Sukop2006,Mohamad2011,Krueger2017}.
It is a discrete particle-based method that is capable to simulate properly the dynamics of complex fluids that flow in complex geometries \cite{Aidun2010}. 
It has been demonstrated that the LBM can recover the solution of many partial differential equations \cite{Wolf-Gladrow2000}, such as the advection-diffusion equation that is the subject of this work.
Different schemes have been proposed to implement boundary conditions for solute concentration within the LBM:
(i) The bounce-back boundary conditions that consist in reflecting back the advected distribution populations that hit a solid boundary. 
A generalized version for mass transport has been proposed by Zhang \textit{et al.}~\cite{Zhang2012},
(ii) Estimating the unknown distribution populations at the boundaries, such as proposed by Inamuro \textit{et al.} \cite{Inamuro2002} for temperature and which can be also used for solute concentration,
(v) Evaluating the unknown physical quantities at curved boundaries using finite difference method techniques, as done by Guo \textit{et al.}~\cite{Guo2015},
(iv) Using a forcing term combined with the immersed boundary method for moving objects, as proposed by Kasparek \textit{et al.}~\cite{Kasparek2016}.
In this work, the forcing term strategy is also used but to include a time-varying jump at zero-thickness membranes separating two domains, as illustrated in Fig.~\ref{fig:mathematical1}.
The mathematical formulation of the problem is given in Sec.~\ref{sec:mathematical}, followed by a brief presentation of the LBM solver in Sec.~\ref{sec:lbm}.
The main steps of the algorithm are detailed in Sec.~\ref{sec:algorithm}, and the parameter calibration explained in Sec.~\ref{sec:calibration}.
Two applications of the algorithm, mass transfer across stationary and moving boundaries, are given in Sec.~\ref{sec:applications}, before concluding the article by Sec.~\ref{sec:conclusions}.
\section{Mathematical formulation}
\label{sec:mathematical}
\begin{figure}
\centering
\includegraphics*[height=1.2in, angle=0]{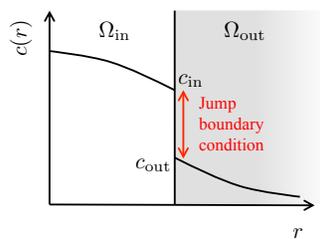}
\caption{\label{fig:mathematical2}
Typical concentration profile $c(r)$, with a jump discontinuity, in the direction normal to a membrane that has a finite resistance. 
}
\end{figure}

Diffusion and advection of a solute in a domain $\Omega$ composed of two subdomains, $\Omega _{\rm in}$ and $\Omega _{\rm out}$, which are separated by a membrane $\partial \Omega$ (see Fig.~\ref{fig:mathematical1}) are described using:
\begin{equation}
\frac{\partial c}{\partial t} + \left( {\bf v} \cdot \nabla c  \right)= D_{\rm in} \nabla ^ 2c \quad \text{if}\quad {\bf r} \in \Omega _{\rm in},
\label{eq:ade_in}
\end{equation}
and
\begin{equation}
\frac{\partial c  }{\partial t} + \left( {\bf v} \cdot \nabla c    \right)= D_{\rm out} \nabla ^ 2c \quad \text{if} \quad {\bf r} \in \Omega _{\rm out},
\label{eq:ade_out}
\end{equation}
where ${\bf r}$ is the vector position and $c({\bf r},t)$ is the local instantaneous solute concentration. 
$D_{\rm in}$ and $D_{\rm out}$ are the diffusion coefficients of the solute in the inner and the outer domains, respectively.
${\bf v}$ is the solvent velocity.
Equations~\ref{eq:ade_in} and \ref{eq:ade_out} are solved, while considering the following boundary condition at the membrane,
\begin{equation}
{\bf J}\cdot{\bf n} = -P \left[ c\right]_{\partial \Omega} = -P \left(c_{\rm out} - c_{\rm in} \right) \quad \text{if} \quad {\bf r} \in \partial \Omega,
\label{eq:bcs}
\end{equation}
that represents the mass flux from $\Omega _{\rm in}$ to $\Omega _{\rm out}$ with a jump in the concentration at the membrane, see Fig.~\ref{fig:mathematical2}. 
$c_{\rm in}$ and $c_{\rm out}$ are the concentrations on the inner and the outer sides of the membrane $\partial \Omega _{\rm in}$ and $\partial \Omega _{\rm out}$, respectively.
$P$ is the permeability of the membrane.
${\bf n}$ is the unit normal vector on the membrane and that is pointing from $\Omega _{\rm in}$ to $\Omega _{\rm out}$.
The main assumptions along this article are:
\begin{itemize}
\item $P$ is constant and uniform along the membrane,
\item The Henry's partition coefficient is unity; therefore, at equilibrium state
\begin{equation}
c_{\rm out}({\bf r},\infty)=c_{\rm in}({\bf r},\infty)=c_{\rm eq}\quad \text{for} \quad {\bf r} \in \partial \Omega,
\end{equation}
is expected, with $c_{\rm eq}$ is the equilibrium concentration,
\item The mass flux is continuous across the membrane: 
\begin{equation}
D_{\rm in} \left( {\bf n} \cdot \nabla c \vert_{\partial \Omega _{\rm in}} \right) = D_{\rm out} \left( {\bf n} \cdot \nabla c \vert_{\partial \Omega _{\rm out}} \right),
\end{equation}
\item The diffusion coefficient is similar in both subdomains: 
\begin{equation}
D = D_{\rm in} = D_{\rm out},
\end{equation}
\item The membrane is solvent-impermeable, which implies the continuity of the normal velocity component across the membrane. 
This explains the absence of the term $c({\bf v} - {\bf v}_{\rm m})\cdot {\bf n}$ on the left hand side of Eq.~\ref{eq:bcs}, when the no-slip boundary condition is considered as well. ${\bf v}_{\rm m}$ is the membrane velocity, 
\item One-way coupling is considered. 
That is the solvent flow alters locally the solute concentration via the advection term $\left( {\bf v} \cdot \nabla c \right)$ in Eqs.~\ref{eq:ade_in} and \ref{eq:ade_out}, while the solute concentration has no impact neither on the solvent flow nor on its hydrodynamical properties. 
This corresponds to the limit of having a scalar field, which is apllicable to dilute solutions,
\item Two-dimensional simulations are performed.
\end{itemize}
For a planar membrane with a stepwise function as the initial condition,
\begin{equation}
c(x,t=0) = \begin{cases}
1 \quad \text{if}\,\, x>0 \\
0 \quad \text{if}\,\,  x<0,
\end{cases}
\end{equation} 
the one-dimensional solution to Eqs.~\ref{eq:ade_in} and \ref{eq:ade_out} along the $x$-direction can be driven analytically, as given by Crank \cite{Crank1975},
\begin{equation}
\begin{split}
c(x,t) & = \frac{1}{2} \left[   1 + \left\lbrace  {\rm erf} \frac{x}{2\sqrt{Dt}}  \right. \right. \\
& + \left. \left. \exp\left(ax + a^2Dt \right) {\rm erfc} \left( \frac{x}{2\sqrt{Dt}} + a\sqrt{Dt}\right) \right\rbrace   \right],
\end{split} 
\label{eq:theo1}
\end{equation}
for $x>0$, and
\begin{equation}
\begin{split}
c(x,t)  & = \frac{1}{2} \left\lbrace  {\rm erfc}\frac{\vert x \vert}{2\sqrt{Dt}} \right. \\
& - \left. \exp \left(a \vert x \vert + a^2 Dt \right) {\rm erfc} \left( \frac{\vert x \vert}{2\sqrt{Dt}} + a\sqrt{Dt}\right) \right\rbrace,
\end{split} 
\label{eq:theo2}
\end{equation}
for $x<0$ with
\begin{equation}
a = 2\frac{P}{D}.
\label{eq:permeability}
\end{equation}
$P$ is the permeability and it is used as a fitting parameter for calibration in Sec.~\ref{sec:calibration}.
\section{Lattice Boltzmann method}
\label{sec:lbm}
The first necessary ingredient for the algorithm is a mass transfer solver. 
In the present work, the LBM is adopted instead of solving directly Eqs.~\ref{eq:ade_in} and \ref{eq:ade_out} using, for example, the finite difference method.
The LBM has emerged as an alternative fluid flow solver in the last decades with high potential to handle complex fluid dynamics, while avoiding the complicated weak formulations and extensive pre-processing meshing needed by the classical finite element method based CFD solvers.
The applications of the LBM have been extended to recover solutions of many partial differential equations, such as the ones of the electromagnetic wave propagation and the advection-diffusion of a solute.
More details about fundamentals and applications of the LBM could be found in the following few available textbooks Refs.~\cite{Wolf-Gladrow2000,Succi2001,Sukop2006,Mohamad2011,Krueger2017}.

The main quantity of interest in the LBM is the distribution function $g_i$.
For the D2Q9 lattice, used here, the two-dimensional discrete position is given by ${\bf r} \equiv (x,y)$ and the nine discrete velocity directions by ${\bf e}_i$, with $i=0\rightarrow8$.
The evolution in time of $g_i$ is given by the lattice-Boltzmann method equation,
\begin{equation}
g_i({\bf r}+{\bf e}_i,t+1) - g_i({\bf r},t) = - \frac{g_i({\bf r},t) - g_i^{\rm eq}({\bf r},t)}{\tau_{\rm d}},
\label{eq:boltzmann}
\end{equation}
where the right-hand side expresses the BGK relaxation operator:
that is $g_i$ relaxes towards its equilibrium distribution $g^{\rm eq}_i$,
\begin{equation}
g^{\rm eq} _{i}({\bf r},t)= \omega _i \, c({\bf r},t) \left[1 + 3({\bf v}\cdot {\bf e}_i) + \frac{9}{2}({\bf v}\cdot{\bf e}_i)^2 -\frac{3}{2}({\bf v})^2 \right],
\label{eq:equilibrium}
\end{equation}
with the D2Q9 lattice weight factors: $\omega _i=4/9$ for $i=0$, $\omega _i=1/9$ for $i=1,\,2,\,3,\,4$ and $\omega _i=1/36$ for $i=5,\,6,\,7,\,8$.
The microscopic relaxation time $\tau_{\rm d}$ in Eq.~\ref{eq:boltzmann} is related to the macroscopic diffusion coefficient via $D = \frac{1}{3}\left( \tau _{\rm d} - \frac{1}{2}\right)$.
In the present work, $\tau_{\rm d}$ is taken within the limited range of $[0.55,1]$, for which the BGK scheme is known to be stable and accurate.
Other collision operators, such as the multi-relaxation time operator (MRT), are used to enhance the stability and the accuracy when using either $\tau_{\rm d} \rightarrow 0.5$ or $\tau_{\rm d} \gg 1$. 
Source terms can be included as forcing terms into the right-hand side of Eq.~\ref{eq:boltzmann}, as is explained below in Sec.~\ref{sec:algorithm} (Eq.~\ref{eq:boltzmann_forcing}).
The local solute concentration can be computed as the zeroth order moment of $g_i$,
\begin{equation}
c(x,y,t) = \sum _{i=0}^{8} g_i(x,y,t).
\label{eq:concentration}
\end{equation}
All the parameters and the physical quantities along this work are given in dimensionless lattice units \cite{Schornbaum2016}.
The spatial and the temporal discretization used in Eq.~\ref{eq:boltzmann} and hereafter are all set to $h =\Delta x = \Delta y = \Delta t = 1$  in dimensionless lattice units, which is the convention adopted in the LBM.
A rigorous physically-based explanation of the lattice units is given by Succi \cite{Succi2001}, and a practical conversion between the lattice units and the physical units (SI) is proposed by Dupin \textit{et al.} \cite{Dupin2007}. 

\section{Algorithm}
\label{sec:algorithm}
The proposed algorithm is useful to handle study cases involving unsteady mass transfer across moving-free membranes with zero thickness and having a finite permeability.
For this, the jump in the concentration could be incorporated within an existing code that uses the LBM to compute the advection and diffusion of a solute.
The jump is introduced by using a forcing term in the LBM equation.
This is somehow similar to the way of introducing hydrodynamic stress jump at deformable moving-free boundaries within the immersed boundary method \cite{Peskin2002}. 
The algorithm is simple and its main consecutive steps need to be implemented right before the streaming step in any existing LBM based code. 

\begin{figure}
\centering
\includegraphics*[height=2.in, angle=0]{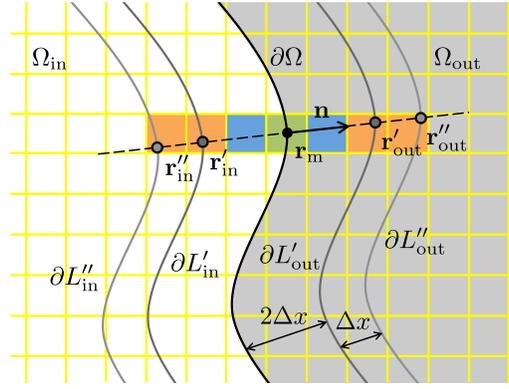}
\caption{\label{fig:algorithm1}
The membrane $\partial \Omega$ and its four closer enveloping layers ($\partial L^{\prime}_{\rm in}$, $\partial L^{\prime\prime}_{\rm in}$, $\partial L^{\prime}_{\rm out}$ and $\partial L^{\prime\prime}_{\rm out}$) with the LBM regular Cartesian grid in the background.
}
\end{figure}

\begin{figure}
\centering
\includegraphics[height=1.15in, angle=0]{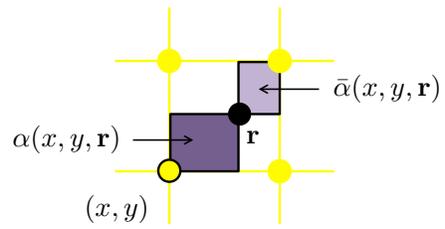}
\caption{\label{fig:algorithm2}
An off-lattice membrane point ${\bf r}$ with two of its areas $\alpha(x,y,{\bf r})$ and $\bar{\alpha}(x,y,{\bf r})$ used respectively for performing the bilinear interpolation and for spreading the source term to the LBM regular Cartesian grid nodes $(x,y)$.
}
\end{figure}

Given the actual membrane position ${\bf r}_{\rm m} \equiv (s,t)$, where $s$ is the curvilinear coordinate along the membrane, and the concentration field $c(x,y,t)$ in the whole computational domain $\Omega$, the forcing term ${\bf Q}(x,y,t)$ needed to compute the concentration for the next time step $c(x,y,t+1)$ is evaluated following the consecutive steps:
\begin{enumerate}
\item Localize the inner and the outer enveloping layers of the membrane,
\begin{equation}
\partial L^{\prime}_{\rm in}, \quad \partial L^{\prime\prime}_{\rm in}, \quad \partial L^{\prime}_{\rm out}\quad \text{and} \quad \partial L^{\prime\prime}_{\rm out},
\end{equation}
whose distances from the actual location of the membrane ${\bf r}_{\rm m}$ are $2h$ and $3h$,  as illustrated and explained in Fig.~\ref{fig:algorithm1},
\item Compute the solute concentration along each layer using the standard bilinear interpolation.
The known concentrations on the four nodes of the orange colored elements in Fig.~\ref{fig:algorithm1} give the concentration, for example, at ${\bf r}^{\prime}_{\rm out}$:
\begin{equation}
c({\bf r}^{\prime}_{\rm out},t) = \sum _{x,y} \alpha(x,y,{\bf r}^{\prime}_{\rm out})c(x,y,t).
\end{equation}
As in Ref.~\cite{Dupin2007}, $\alpha$ is the area shared by an off-lattice point ${\bf r}$ and a nearby on-lattice node $(x,y)$ of the LBM regular Cartesian grid, see Fig.~\ref{fig:algorithm2}.

Layers that are $1h$ distant from the membrane are avoided because the concentration there is not necessary continuous and thus not practical for interpolation, as is the case for the blue colored elements in Fig.~\ref{fig:algorithm1}.
The smooth Diract function usually used in the immersed boundary method \cite{Peskin2002} is not adapted here for holding the sharp discontinuity in the concentration at the membrane.
Instead, it does smear out numerically the concentration if used,
\item Compute the concentrations on the inner and the outer sides of the membrane using the linear extrapolation,
\begin{align}
c_{\rm in}(s,t) & = \left[3c({\bf r}_{\rm  in}^{\prime},t) - 2c({\bf r}_{\rm in}^{\prime\prime},t)\right]h, \\
c_{\rm out}(s,t) & = \left[3c({\bf r}_{\rm out}^{\prime},t) - 2c({\bf r}_{\rm out}^{\prime\prime},t)\right]h.
\end{align}
These give an estimation of the jump in the concentration at the membrane,
\item Compute the \textit{ad hoc} source term ${\bf q}(s,t)$ along the membrane by multiplying the above jump by a numerical parameter $R$, hereafter called resistance,
\begin{equation}
{\bf q}(s,t) = -R \left[c_{\rm out}(s,t) - c_{\rm in}(s,t) \right]{\bf n}.
\end{equation}
This has non-zero value only on the membrane location and zero elsewhere,
\item Spread the source term to the overall computational domain by evaluating,
\begin{equation}
{\bf Q}(x,y,t) = \int _{\partial \Omega} \bar{\alpha}(x,y,s){\bf q}(s,t) {\rm d}s,
\end{equation}
on the nodes $(x,y)$ surrounding closely each membrane node ${\bf r}_{\rm m}$, 
\item Plug the source term into the LBM equation as a forcing term:
\begin{equation}
\begin{split}
g_i({\bf r}+{\bf e}_i,t+1) - g_i({\bf r},t) &= - \frac{g_i({\bf r},t) - g_i^{\rm eq}({\bf r},t)}{\tau_{\rm d}} \\
& + \frac{\omega _i}{c^2_{s}}\left( {\bf Q}(x,y,t) \cdot {\bf n} \right),
\end{split}
\label{eq:boltzmann_forcing}
\end{equation}
in such way ${\bf Q}(x,y,t) \cdot {\bf n}$ controls the mass flux across the membrane.
\end{enumerate}

The algorithm recovers the solution of Eqs.~\ref{eq:ade_in} and \ref{eq:ade_out} with the boundary condition Eq.~\ref{eq:bcs}, as well as any other advection diffusion equation of the type:
\begin{equation}
\frac{\partial c  }{\partial t} + \left( {\bf v} \cdot \nabla c    \right)=  \nabla \cdot \left( D \nabla c \right)+ q\delta
\label{eq:ade}
\end{equation}
that is valid throughout the whole computational domain $\Omega$, without distinction between its subdomains.
$q$ is a source term with $\delta$ a function that takes unity as a value on the membrane and zero elsewhere.

\begin{figure}
\centering
\subfloat[]{\label{fig:algorithm3a}\includegraphics[height=1.7in, angle=0]{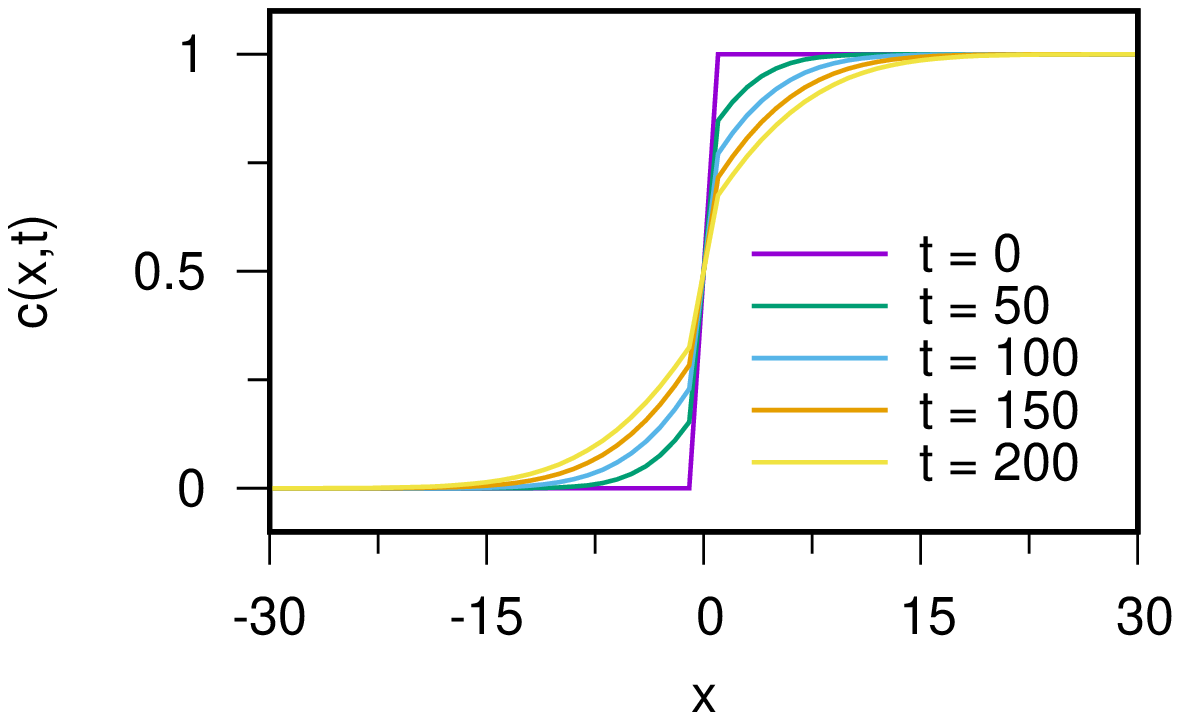}}
\quad\
\subfloat[]{\label{fig:algorithm3b}\includegraphics[height=1.7in, angle=0]{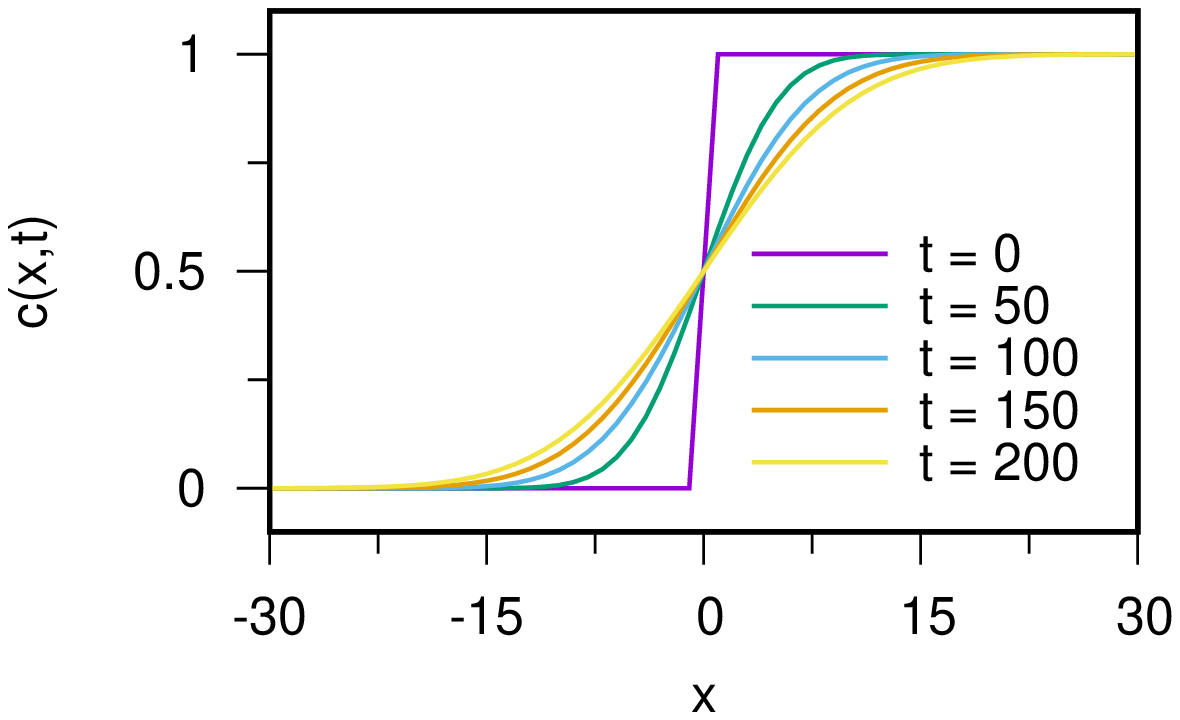}}
\caption{\label{fig:algorithm3}
Evolution in time of numerically computed concentration profiles $c(x,t)$, along the $x$-direction, for a planar membrane that has a finite resistance $R=0.15$ (a) For comparison purpose, profiles obtained for $R=0$ are also shown in (b).
}
\end{figure}

Figure~\ref{fig:algorithm3a} gives the obtained concentration profiles over time for a planar membrane located at $x_{\rm m}=0$ when using the algorithm with an arbitrary non-zero value of the resistance $R=0.15$. 
The solute concentration profiles are steep and demonstrate a slow diffusion process in Fig.~\ref{fig:algorithm3a} compared to the case with zero resistance $R=0$, that is with an infinitely permeable membrane ($P \rightarrow \infty$), shown in Fig.~\ref{fig:algorithm3b}.
The algorithm produces successfully and surprisingly the expected qualitative behavior of the solutions of Eq.~\ref{eq:ade_in} and Eq.~\ref{eq:ade_out} with the unsteady jump boundary condition at the membrane Eq.~\ref{eq:bcs}.
\section{Calibration} 
\label{sec:calibration}
The delicate non-trivial step in the algorithm is how to set the adequate numerical value for the resistance $R$ (Step $4$ in Sec.~\ref{sec:algorithm}).
The algorithm needs a value that reproduces faithfully the same mass transport scenario that a membrane with a given known permeability $P$ does.
Running a series of trailer and error tests to calibrate the computer code can solve this issue.
The exact one-dimensional analytical solution, Eqs.~\ref{eq:theo1} and \ref{eq:theo2}, is used as a reference.
In each calibration test, a numerical value is attributed to $R$ and the resulting computed concentration profiles are fit with the analytical solution.
The permeability $P$ in Eq.~\ref{eq:permeability} is used as a fitting parameter and it is tuned until accomplishing excellent fit with the numerical data.

In this section, a planar membrane is presented by a discretized straight line along the $y$-direction, whose nodes are uniformly separated by $h=1$ and are set on $x_{\rm m}=0$ along the $x$-direction and on non-zero along the $y$-direction ($y_{\rm m} \neq 0$).
Periodic boundary conditions are set along the top and the bottom edges of the computational domain in order to mimic an infinitely symmetric system in the $y$-direction, and thus to recover the one-dimensional solution.
Zero-flux is set on the two extreme right and left edges of the simulation domain in the $x$-direction, i.e. at $x=-100$ and $x=100$.

Computed numerical concentration profiles over time $t$ for a non-zero value of the resistance $R=0.16$, with their respective fit, are reported in Figs.~\ref{fig:calibration1a} and \ref{fig:calibration1azoom}.
The fitting parameter $P$ is tuned until reaching a good fit.
The numerical data are presented by symbol points, while the analytical solution with solid lines.
It is remarkable how a single value $P$ can fit reasonably the solutions taken at any given time.
Figure~\ref{fig:calibration1b} shows the instantaneous local relative error computed as the deviation of the computed solution from the theory, 
\begin{equation}
{\rm Error}(x,t) = 100\times  \abs[\Big]{ \frac{c_{\rm num}(x,t) - c_{\rm theo}(x,t)}{c_{\rm theo}(x,t)}}
\end{equation}
where $c_{\rm num}(x,t)$ is the numerically computed local concentration and $c_{\rm theo}(x,t)$ the expected theoretical value (Eqs.~\ref{eq:theo1} and \ref{eq:theo2}).
Because of the antisymmetric nature of the concentration profile with respect to the point ($x=0$,$c=0.5$), the error is computed and reported only for $x>0$.
Along the overall simulation, each local computed error is less than $10\%$; therefore, the accuracy of the proposed numerical method can be qualified as good.
Figure~\ref{fig:calibration1c} gives the adequate value of $P$ found for each imposed value of $R$.
The possible maximum value of the resistance that can be imposed is $R=0.16$ for $\tau _{\rm d} = 1$.
Beyond this value the code produces unphysical results.
However, its minimum values can of course goes down to zero, which corresponds to an infinitely permeable membrane ($P\rightarrow \infty$).
At first glance, $P$ seems to be inversely proportional to $R$, but it is not the case because the proportionality coefficient varies with $R$.
Another important detail is that $R$ must be updated whenever the diffusivity of the subdomains $D$ is changed.
For example, when having $R=0.16$ for $\tau _{\rm d} = 1$, the value of $R$ needs to be varied as shown in the inset of Fig.~\ref{fig:calibration1c}.
It needs to be decreased whenever $\tau _{\rm d}$ is decreased.
However, it becomes delicate to evaluate its value when $\tau _{\rm d}$ tends to $1/2$ because of the numerical instabilities.

Another factor is found to have also an effect on the computed solution.
It is the location of the membrane.
If the $x$-coordinate of all the membrane nodes $x_{\rm m}$ are off-lattice then the fit is not anymore good as if they are on-lattice, see Fig.~\ref{fig:calibration2}. 
The code still produces data that are reasonably comparable to the expected theoretical solution, but with a slight deviation.
However, the jump character of the concentration at the membrane is still recovered.
For the position of the membrane nodes in the $y$-direction, it is found that it does not have any influence.
Exactly the same concentration profile is computed whether $y_{\rm m}$ is on- or off-lattice.
\section{Applications}
\label{sec:applications}
An LBM based code is used together with the proposed algorithm to carry out two-dimensional simulations of solute controlled release from particles.
The particles are initially loaded inside with a given initial concentration of a solute $c_0$ that later on diffuses across the membrane towards the external surrounding environment $\Omega _{\rm out}$.
With the proposed algorithm, the release rate is controlled by tuning the membrane permeability $P$ whose adequate corresponding resistance $R$ is picked up from the calibration in Fig.~\ref{fig:calibration1c}.
The reported data are not meant to recover or to predict quantitatively any experimental data, but rather to demonstrate briefly the effectiveness of the algorithm to reproduce the sharp jump in the solute concentration at the membrane.
\subsection{Solute release from a stationary particle}
A circular particle with radius $R_{\rm p}=20$ is placed at the center of a bounded square domain of size $160\times160$.
Zero-flux boundary conditions are set along the four bounding edges of the square using the bounce-back boundary condition \cite{Sukop2006}.
The initial concentration $c(x,y,t=0)$ is set to a non-zero value $c_{0}=1$ inside the particle and zero elsewhere.
This situation mimics a reservoir that is loaded initially with a solute that diffuses later on across the membrane to the outer surrounding domain, while the solvent is at rest and the particle does not undergo any shape deformation (${\bf v}({\bf r},t)=0$ for $\forall {\bf r}\in \Omega$ and $t\in[0,\infty[$).
Two extreme situations are modeled: 
(i) a particle with finite permeability $R=0.16$ and 
(i) a particle with infinite permeability that corresponds to $R=0$.

Figures~\ref{fig:fig5a} and \ref{fig:fig5b} give the spatial distribution of the concentration $c(x,y,t)$ in the whole computational domain at time $t=300$.
For both cases, the diffusion is isotropic and symmetric with respect to the center of the particle.
However, inside the cylinder is more reddish in Fig.~\ref{fig:fig5a} than is the case in Fig.~\ref{fig:fig5b}.
For the finite permeability case, the solute diffusion across the membrane is remarkably restricted; therefore, more solute is retained inside the particle.
This difference is clearly noticeable when reporting the solute concentration profiles at the same time and along the radial coordinate $r$, as shown in Fig.~\ref{fig:fig5c}.
The concentration shows a smooth variation for the case of infinitely permeable particle, while it exhibits a very steep jump at the membrane for the restricted permeability case.

The restrictive permeability of the particle has an effect on another interesting global physical quantity that is the overall release of the encapsulated solute, which is reported in Fig.~\ref{fig:fig5d}.
It is computed as,
\begin{equation}
M(t) = 100 \times \left[ \frac{m(0) - m(t)}{m(0)}\right],
\label{eq:release}
\end{equation}
where $m(t)$ is the amount of the remaining solute mass inside the cylinder at time $t$:
\begin{equation}
m(t) = \int _{\Omega _{\rm in}} c(x,y,t){\rm d}x{\rm d}y,
\label{eq:mass}
\end{equation}
A finite permeability slows down dramatically the amount of the released solute.
However, the two curves in Fig~\ref{fig:fig5d} converge into a single one at long time when the system and the concentration on both sides of the membrane tend to equilibrium, 
\begin{equation}
c_{\rm in}(s,\infty) = c_{\rm out}(s,\infty) = c_{\rm eq}.
\label{eq:eq_concentration}
\end{equation}
This later leads to a vanishing forcing term and thus the resistance effect comes off.
The impact of the restricted permeability is very noticeable at short time before the system reaches equilibrium.
The resulting discontinuity jump in the concentration at the membrane is unsteady and evolves in time by decaying when the concentration on both sides of the membrane come closer to their equal equilibrium value Eq.~\ref{eq:eq_concentration}.

The above case study demonstrates the capability of the algorithm to reproduce qualitatively the jump discontinuity in the solute concentration at the membrane, and its expected unsteady decay over time.
To the best of the author's knowledge and as also stated by Sherk \cite{Sherk2011}, no exact analytical solution exists for the above case study that would allow to perform a precise quantitative accuracy study.
Nevertheless, the proposed algorithm reproduces qualitatively the same results as computed by both Huang \textit{et al.} \cite{Huang2009} and Sherk \cite{Sherk2011} using the IBM.
\subsection{Solute release from a flowing fluid-filled particle}
\label{sec:deformable}
The computer code, with the proposed algorithm, can also be applied to another class of problems that deal with fluid-solute-structure interaction.
For example, oxygen uptake and delivery by red blood cells or drug delivery by vesicles.
Here, the algorithm is used to handle the jump boundary conditions in the solute concentration at a moving deformable membrane of a fluid-filled particle flowing in a channel.
The interactions between different elements of such system is summarized in Fig.~\ref{fig:fssi}.
The flow transports and deforms the membrane, which in its turn disturbs back the flow at its vicinity.
The membrane is assumed to be inextensible and resists to bending.
All the details concerning the fluid-structure two-way coupling via the IBM and the mechanics of the particle membrane have already been reported by the same author, see Refs.~\cite{Kaoui2011,Kaoui2012,Kaoui2014,Kaoui2016}.
The flow is also computed by the LBM, as done in Refs.~\cite{Kaoui2017,Kaoui2018}.

The originality of the present work consists in implementing an explicit coupling between the membrane dynamics and the restrictive solute transport across it.
In contrast to Refs.~\cite{Jayathilake2010,Jayathilake2011,Miyauchi2015,Miyauchi2017,Yao2017}, where the particle membrane is permeable to both solute and solvent, here the membrane is perfectly solvent-impermeable.
It allows only solute mass transport with finite permeability.
This is a accomplished numerically via the proposed algorithm (Sec.~\ref{sec:algorithm}).
In two recent works  by the same author \cite{Kaoui2018} and another by Kabacao\v{g}lu et \textit{al.} \cite{Kabacaoglu2017}, only one-way coupling is reported where the flow advects the solute but no explicit mass transfer boundary conditions for the solute at the membrane has been implemented.
This is limited to situations of having membranes with extremely large permeability \cite{Fournier1999}.

In the present application, two extreme cases are again performed to appreciate the impact of including a resistance to the membrane permeability.
For these simulations, the computational domain is $400\times100$, the maximum solvent flow velocity at the channel centerline is $v_{\rm max}= 0.084$, the solvent viscosity is $\nu = 0.167$ ($\tau _{\rm f}=1$), the diffusion coefficient is $D=0.017$ ($\tau _{\rm d} = 0.55$), and the particle effective radius is $R_{\rm p}=20$.
The membrane is discretized into $120$ points uniformly separated by $\Delta s=h=1$, which is suitable for the IBM. 
The initial concentration $c(x,y,t=0)$ is $c_{0}=1$ inside the particle and zero elsewhere.
The particle has initially an elliptical shape and it is initially placed at the channel centerline.
Later on it moves under flow and deforms, while conserving its enclosed area and its membrane perimeter.
\\
The two snapshots in Figs~\ref{fig:liposome1} and \ref{fig:liposome2} show the distribution of the solute in the whole computational domain.
The flow direction is upwards in both figures.
The particle moves while releasing its encapsulated solute and undergoing shape deformation.
Again it is appreciable how the non-zero resistance slows down the diffusion of the solute from inside the particle to its surrounding suspending flowing fluid.
Inside the particle is reddish in Fig.~\ref{fig:liposome1} compared to the Fig.~\ref{fig:liposome2}.
This outcome can be also appreciated in the solute concentration profiles along the $x$-direction reported in Fig.~\ref{fig:liposome3}, and that shows the jump in the concentration at the membrane for the case with non-zero resistance $R=0.0028$, in particular at the front of the particle.
Figure~\ref{fig:liposome_release} gives the release rate of the solute from the particle for the two extreme cases, with and without resistance.
A non-zero value of the resistance restricts the solute transport across the membrane, which slows down the overall release.

Again this case study demonstrates the effectiveness of the proposed algorithm to reproduce faithfully the jump in the concentration at membranes of even moving deformable particles.
There is no available comparable data as shown in Fig.~\ref{fig:liposome} obtained by other theoretical, numerical or experimental methods for qualitative or quantitative validation.
The application situations computed in Refs.~\cite{Jayathilake2011,Gong2014,Yao2017} are for membranes with different mechanics and different type of mass transfer boundary conditions at the membrane.
Here, the solute transport across the membrane takes place only by pure diffusion mechanism.
\subsection{Non-permeable membranes}
\label{sec:nonpermeable}
The proposed algorithm can also handle the case of strictly non-permeable membranes by setting a constant non-evolving forcing term, as demonstrated in Fig.~\ref{fig:nonpermeable} for ${\bf q}(s,t)= -0.17\,{\bf n}$.
This corresponds to the situation of having a non-permeable membrane that does not allow any exchange of solute.
No diffusion takes place across the membrane; therefore, the initial step-shaped concentration profile is held unchanged in time.
Other parameters of the simulation: $\tau _{\rm d} = 1$ and the rest is the same as used in Sec.~\ref{sec:calibration}.
\section{Concluding remarks}
\label{sec:conclusions}
The proposed algorithm is simple to implement in an existing lattice Boltzmann method based code. 
It uses an \textit{ad hoc} forcing term that surprisingly capture properly the expected behavior of solute transport across membranes with restricted permeability, and even for non-permeable membranes.
It recovers the unsteady sharp jump in the concentration at the membrane and that decays over time when the concentration on both sides of the membrane approaches its equilibrium value.
The computer code including the algorithm is validated against the exact one-dimensional analytical solution given by Crank \cite{Crank1975} for the planar membrane case.
The instantaneous concentration profiles are perfectly computed for membranes with nodes located exactly on-lattice and reasonably for off-lattice positions.
Moreover, the algorithm can also handle problems involving fluid-solute-structure interaction and particles with arbitrary smooth shapes, in particular deformable membranes.
These features have been demonstrated by two case studies: mass transfer from a stationary rigid particle and from a moving deformable fluid-filled particle.
The obtained results are qualitatively comparable to the ones reported in Refs.~\cite{Huang2009,Sherk2011} for the case of a circular particle at rest.
However, comparison with previous studies performed with the immersed interface method is not possible because in the present work the membrane is perfectly solvent-impermeable.
Extension to three-dimensional space and adaptation to heat transfer are expected to be straightforward.

\begin{figure}
\centering
\subfloat[]{\label{fig:calibration1a}\includegraphics[height=2.in, angle=0]{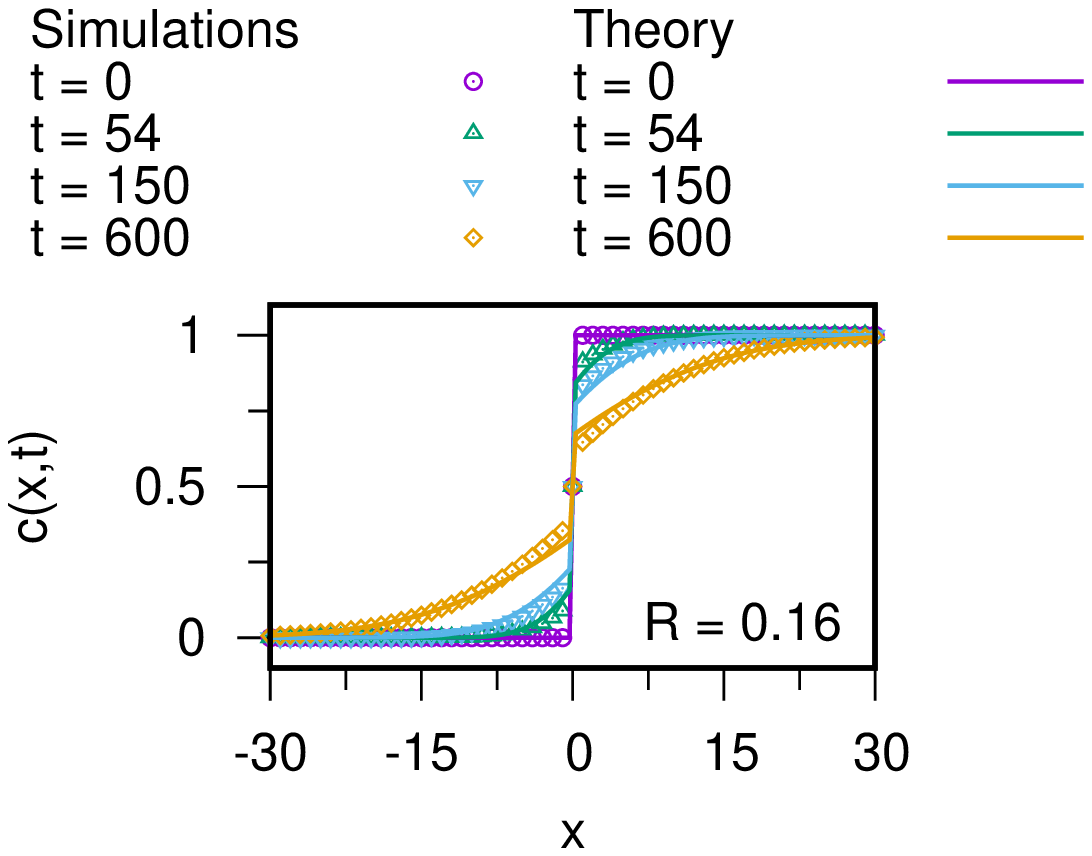}}
\subfloat[]{\label{fig:calibration1azoom}\includegraphics[height=1.55in, angle=0]{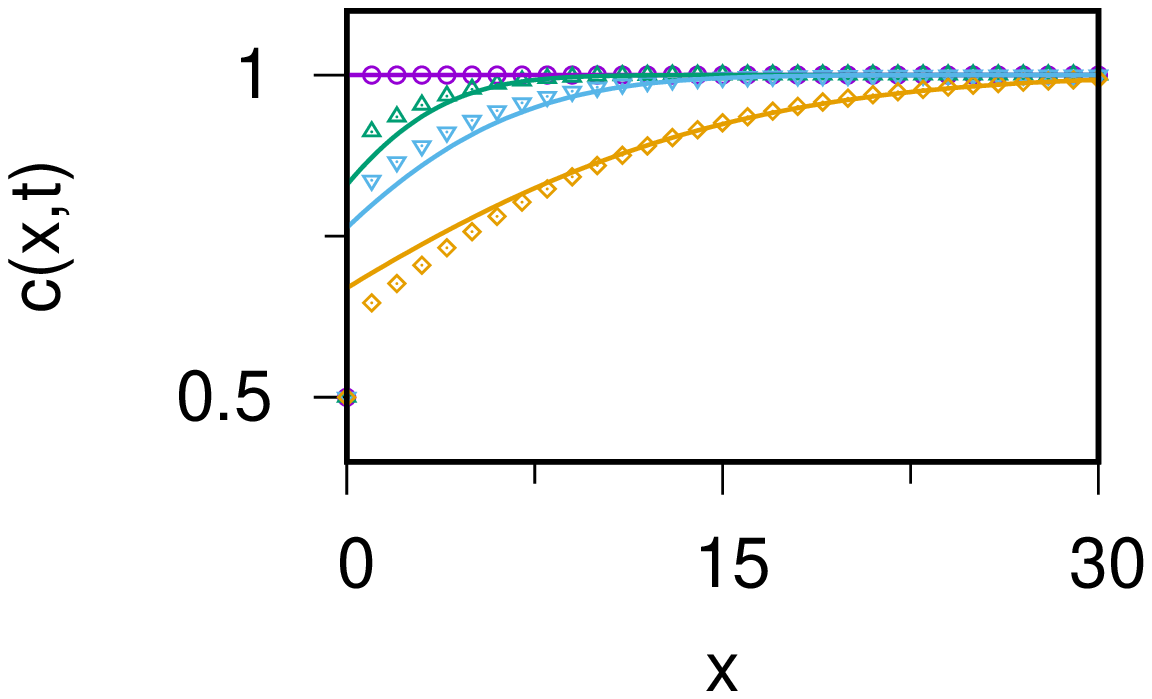}}
\\
\quad\quad\quad\subfloat[]{\label{fig:calibration1b}\includegraphics[height=1.8in, angle=0]{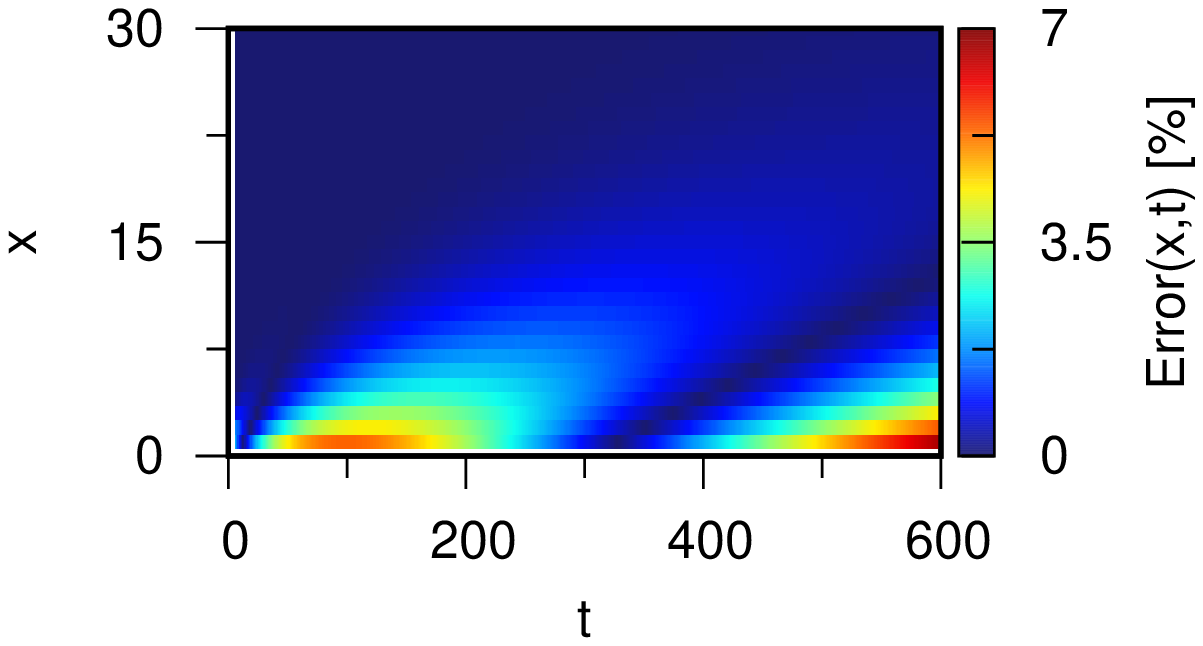}}
\\
\subfloat[]{\label{fig:calibration1c}\includegraphics[height=1.8in, angle=0]{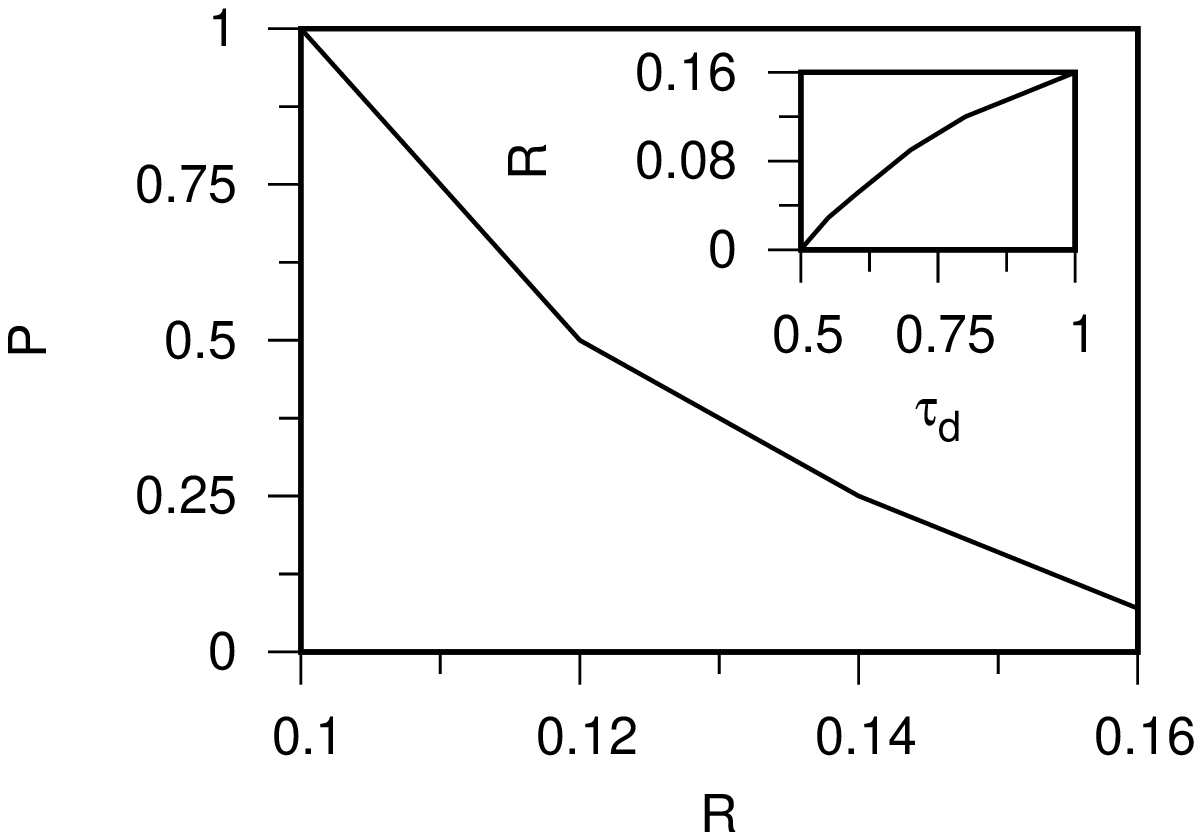}}
\caption{\label{fig:calibration1}
(a) Fit of numerically computed concentration profiles $c(x,t)$, at different times $t$, with the one-dimensional analytical solution,
(b) Zoom into the region of $x>0$,
(c) Evolution in time of the local error ${\rm Error}(x,t)$ of the computer code,
(d) The adequate fitting parameter $P$ for each resistance $R$.
The inset gives how $R$ should be adapted when the diffusivity $D$ is changed.
}
\end{figure}

\begin{figure}
\centering
\subfloat[]{\label{fig:calibration2a}\includegraphics[height=1.75in, angle=0]{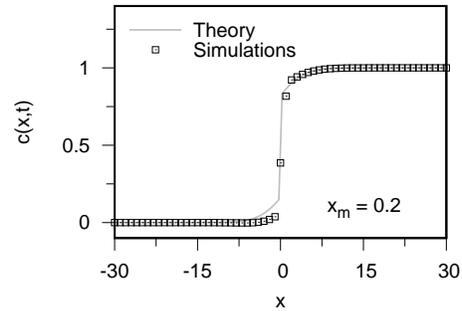}}
\\
\subfloat[]{\label{fig:calibration2b}\includegraphics[height=1.75in, angle=0]{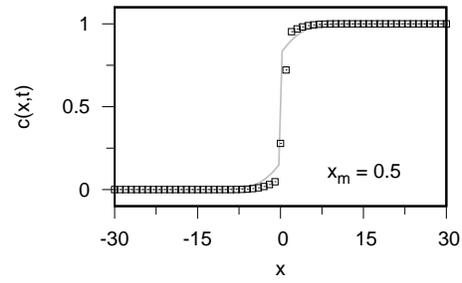}}
\\
\subfloat[]{\label{fig:calibration2c}\includegraphics[height=1.75in, angle=0]{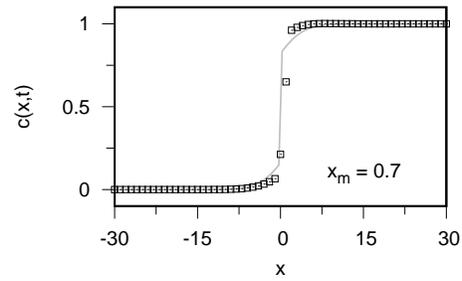}}
\caption{\label{fig:calibration2}
Numerically computed concentration profiles $c(x,t)$, and the expected analytical solution, at time $t=54$ and for three arbitrary chosen off-lattice positions of a planar membrane: $x_{\rm m} = 0.2$, $x_{\rm m} = 0.5$ and (a) $x_{\rm m} = 0.7$. The resistance is $R=0.16$.
}
\end{figure}

\begin{figure}
\centering
\subfloat[]{\label{fig:fig5a}\includegraphics*[height=1.7in, angle=0]{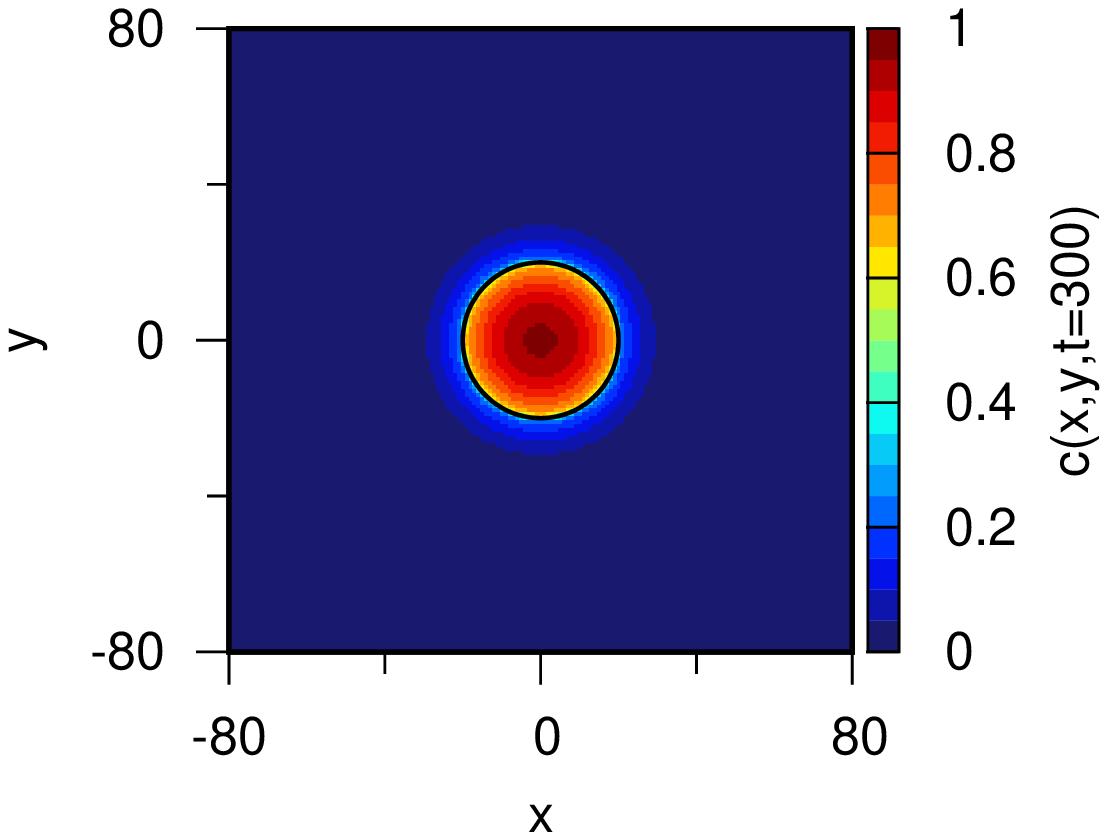}}
\quad
\subfloat[]{\label{fig:fig5b}\includegraphics*[height=1.7in, angle=0]{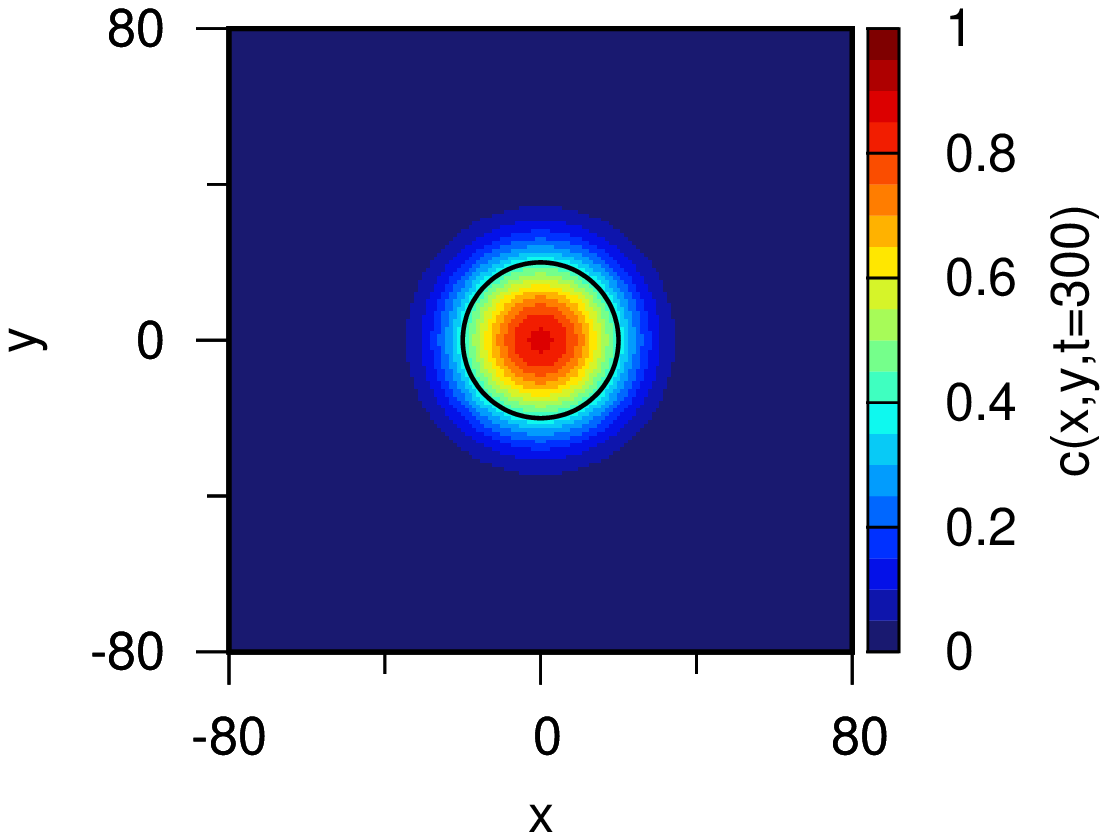}}
\\
\subfloat[]{\label{fig:fig5c}\includegraphics*[height=1.7in, angle=0]{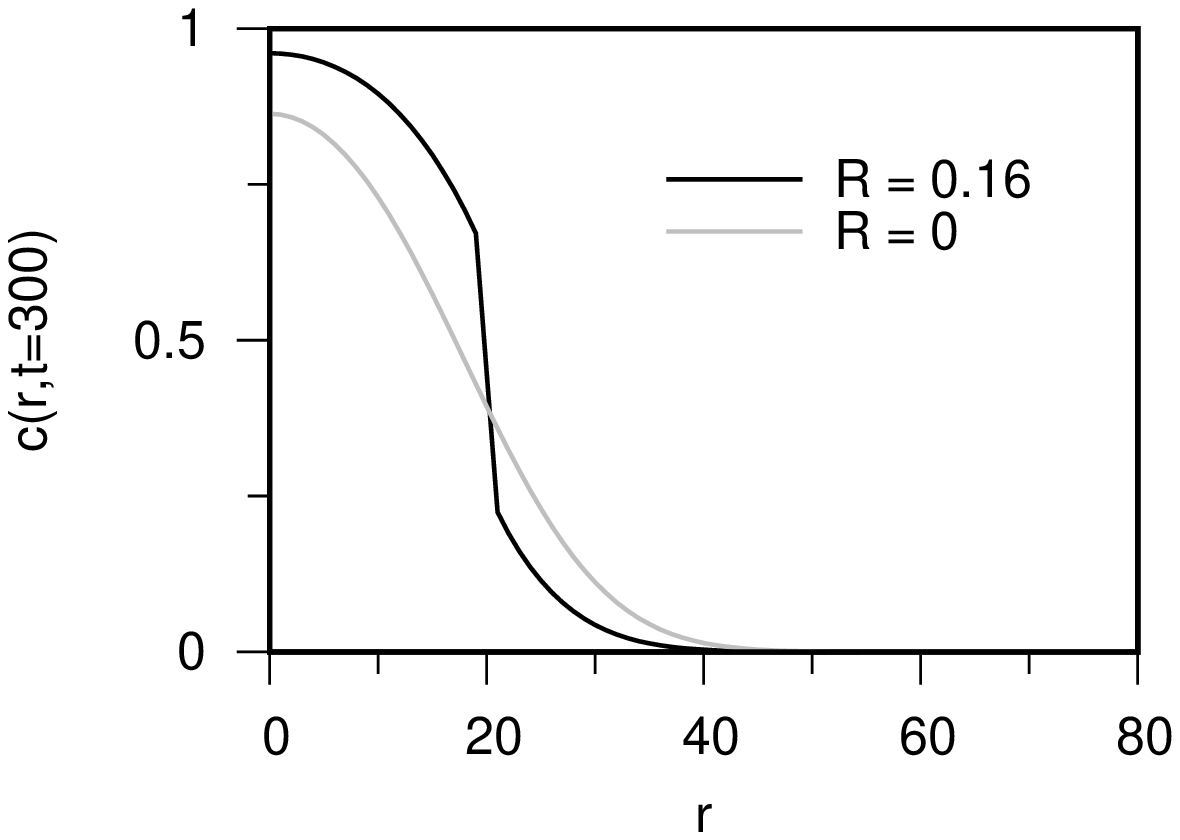}}
\\
\subfloat[]{\label{fig:fig5d}\includegraphics*[height=1.7in, angle=0]{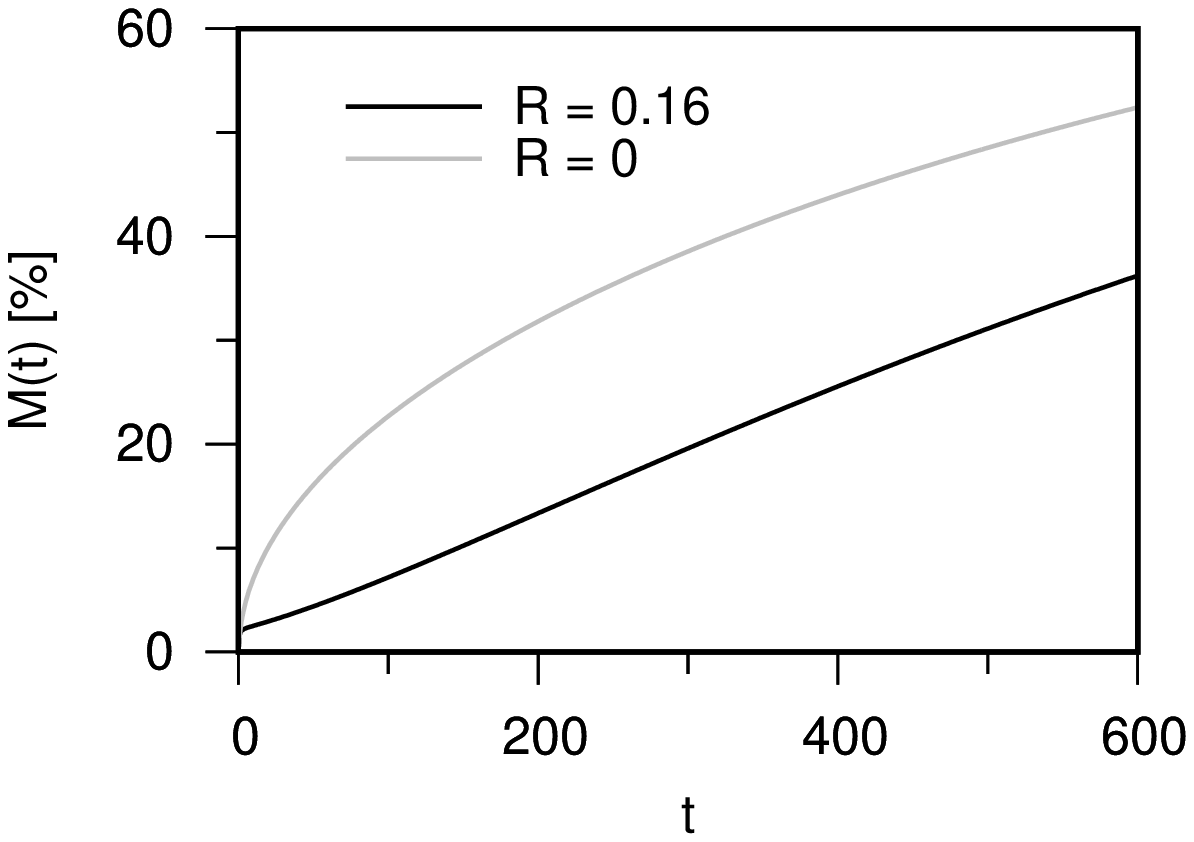}}
\caption{\label{fig:fig5}
Comparison between the spatial distribution of the solute concentration $c(x,y,t)$ at time $t=300$ for a stationary rigid particle with finite permeability $R=0.16$ (a) and an infinite permeability $R=0$ (b), the contour in black solid line represents the membrane of the particle. (c) Numerically computed concentration profiles along the radial coordinate $r$ and (d) Evolution in time of the solute release $M(t)$ before the system reaches the equilibrium state.
}
\end{figure}

\begin{figure}
\centering
\includegraphics*[height=1.in, angle=0]{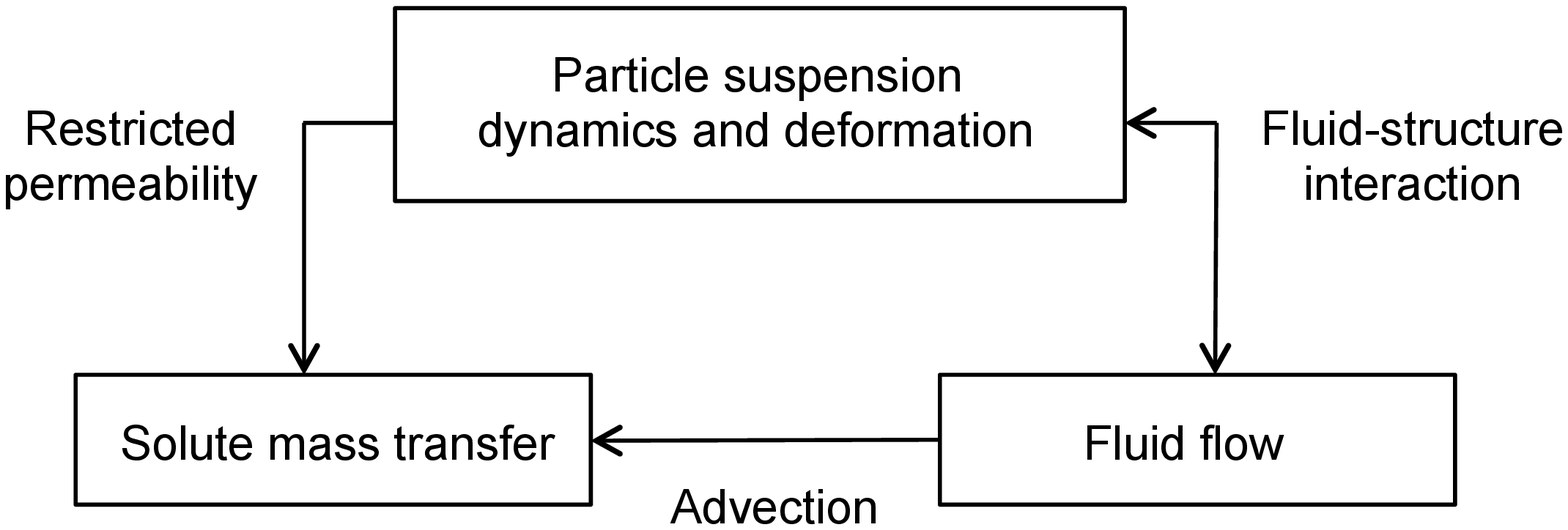}
\caption{\label{fig:fssi}
Interactions between different elements of a problem involving fluid-solute-structure interaction, which is studied in Sec.~\ref{sec:deformable}.
}
\end{figure}

\begin{figure}
\centering
\subfloat[]{\label{fig:liposome1}\includegraphics[height=1.7in, angle=0]{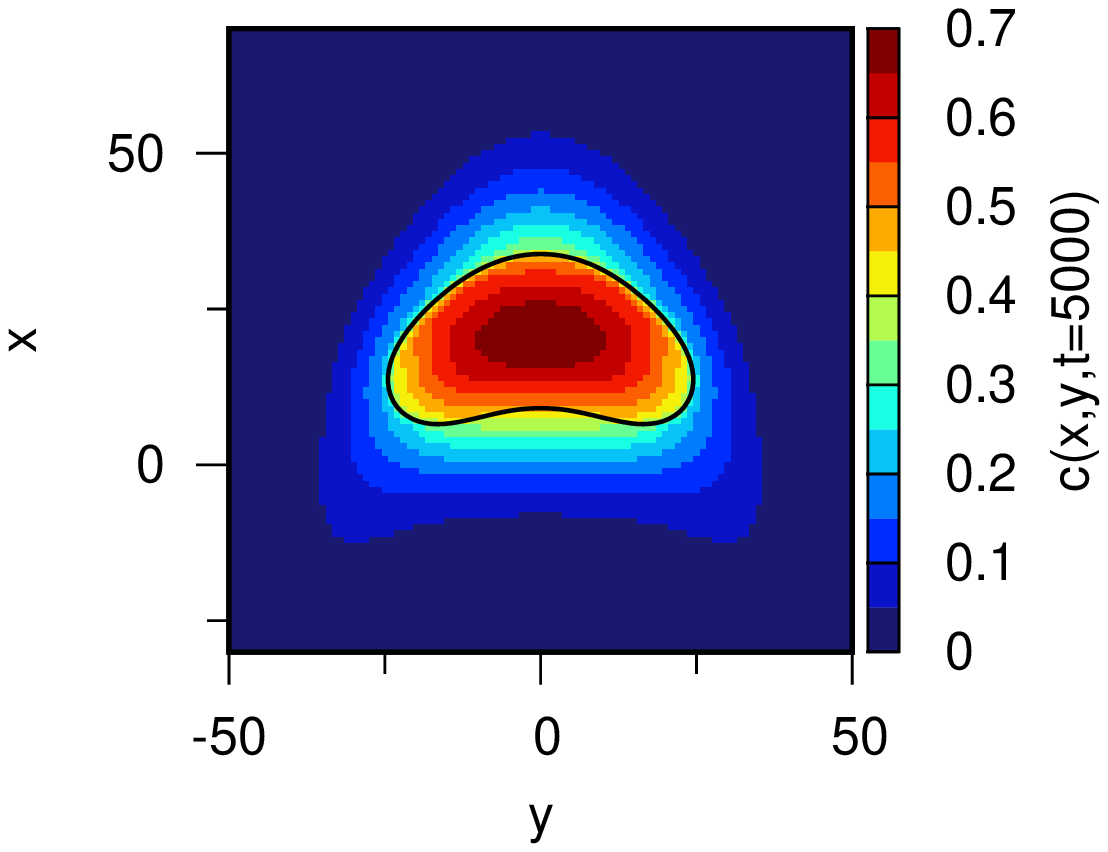}}
\quad
\subfloat[]{\label{fig:liposome2}\includegraphics[height=1.7in, angle=0]{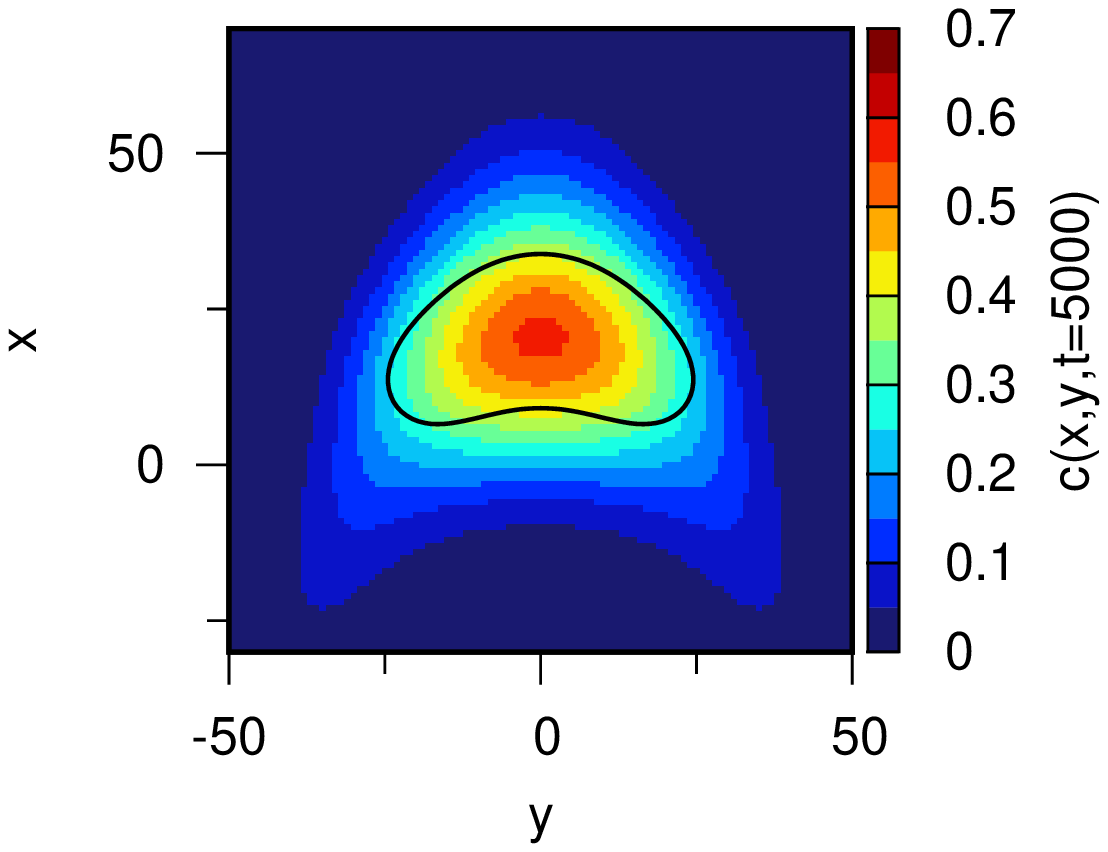}}
\\
\subfloat[]{\label{fig:liposome3}\includegraphics[height=1.7in, angle=0]{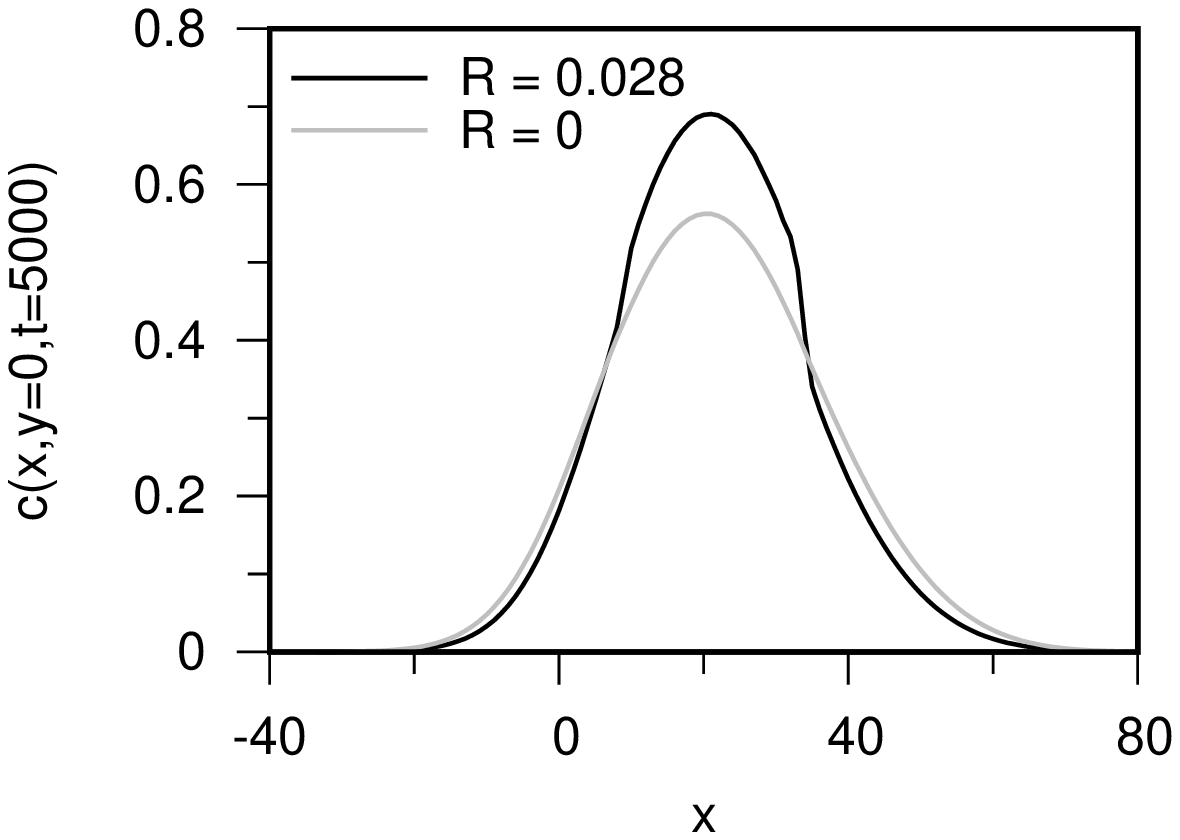}}
\\
\subfloat[]{\label{fig:liposome_release}\includegraphics[height=1.7in, angle=0]{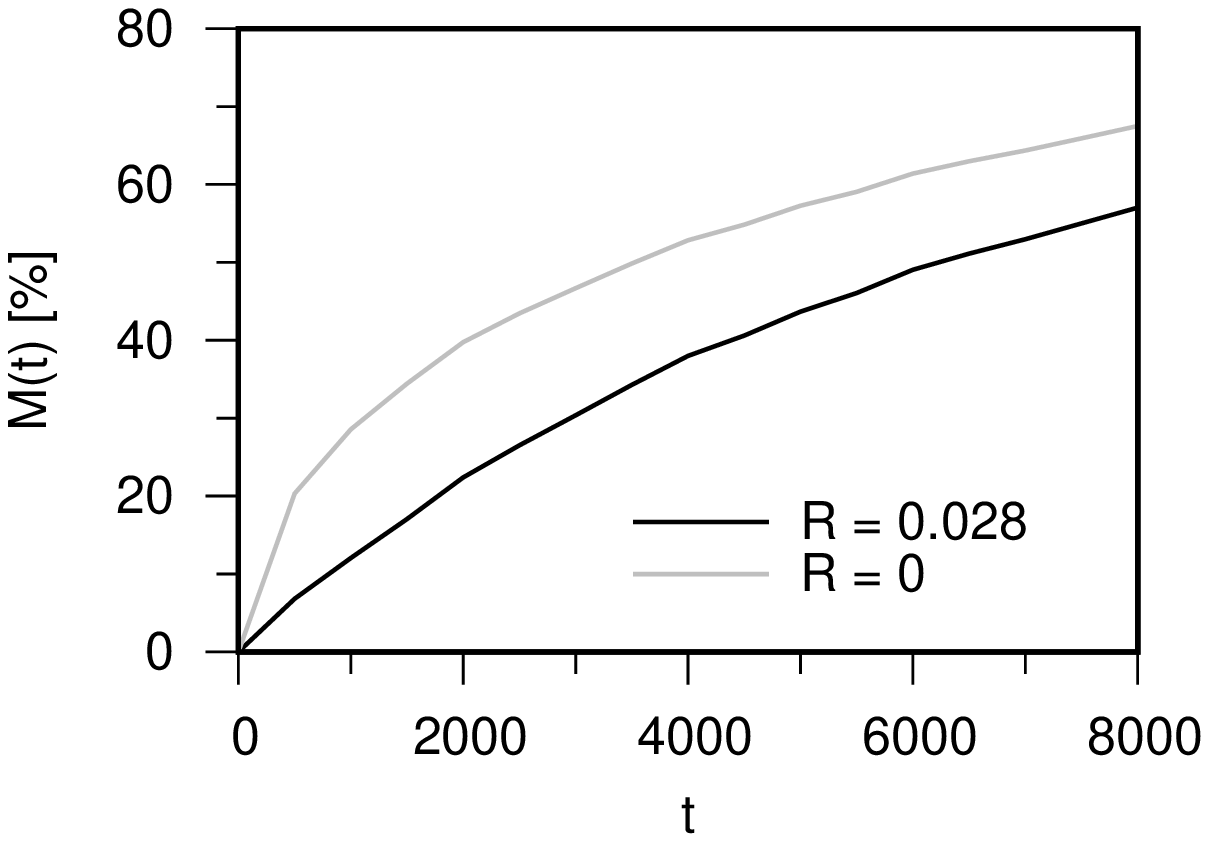}}
\caption{\label{fig:liposome}
The spatial distribution of the solute concentration inside and around a fluid-filled deformable particle flowing in a channel at $t=5000$:
(a) a particle with a membrane presenting a resistance $R = 0.028$, and
(b) a particle with an infinitely permeable membrane $R=0$,
the contour in black solid line represents the membrane of the fluid-filled particle,
(c) The solute concentration profiles along the $x$-direction,
(d) The solute release rate in time $M(t)$ for the two cases.
}
\end{figure}

\begin{figure}[p!]
\centering
\includegraphics[height=1.7in, angle=0]{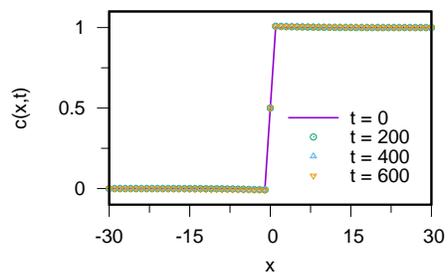}
\caption{\label{fig:nonpermeable}
Computed concentration profiles, at different times, for a non-permeable stationary planar membrane achieved by setting a constant source term.
}
\end{figure}

The major open issues with the proposed algorithm are the physical meaning and the mathematical derivation of the \textit{ad hoc} introduced numerical parameter $R$, which are missing and need to be elucidated.
$R$ can be seen as the membrane resistance, which is usually defined as the inverse of the permeability $\frac{1}{P}$ (or $\frac{\delta _{\rm m}}{P}$ for membranes with non-zero thickness $\delta _{\rm m}$).
However, it is not the case here as shown in Fig.~\ref{fig:calibration1c}.
The proportionality coefficient seems to depend on other parameters, for example, the diffusivity as shown in the inset of Fig.~\ref{fig:calibration1c}.
Moreover, the convergence analysis in the LBM is usually achieved by rescaling the parameters and the physical quantities instead of refining grid resolution and reducing the time step.
However, this has not been possible because the rescaling is not obvious when the resistance is non-zero.
This is under investigation and it will be the subject of a future publication. 
Nevertheless, the computer code with the proposed algorithm and the \textit{ad hoc} incorporation of the forcing term captures the essential physics of solute transport across stationary and moving deformable membranes with restricted permeability.
\newpage
\clearpage
\clearpage
\clearpage
\clearpage
\clearpage
\clearpage
\clearpage
\clearpage
\clearpage
\clearpage
%
\end{document}